\documentclass[prd,aps,nofootinbib,superscriptaddress,showpacs,floatfix,preprintnumbers]{revtex4}
\usepackage{hyperref,amssymb,amsmath,graphicx,xcolor}

\def \non {\nonumber}
\def \beq {\begin{equation}}
\def \eeq {\end{equation}}

\begin{document}

\title{Pion Distribution Amplitude from Lattice QCD}

\author{Jian-Hui Zhang}
\email{jianhui.zhang@ur.de}
\affiliation{Institut f\"ur Theoretische Physik, Universit\"at Regensburg, 
D-93040 Regensburg, Germany}

\author{Jiunn-Wei Chen}
\email{jwc@phys.ntu.edu.tw}
\affiliation{Department of Physics, Center for Theoretical Sciences, and Leung Center for Cosmology and Particle Astrophysics, National Taiwan University, Taipei, Taiwan 106}
\affiliation{Center for Theoretical Physics, Massachusetts Institute of Technology, Cambridge, MA 02139, USA}

\author{Xiangdong Ji}
\email{xji@umd.edu}
\affiliation{Tsung-Dao Lee Institute, and College of Physics and Astronomy, Shanghai Jiao Tong University, Shanghai, 200240, P. R. China}
\affiliation{Maryland Center for Fundamental Physics, Department of Physics, University of Maryland, College Park, Maryland 20742, USA}

\author{Luchang Jin}
\email{ljin.luchang@gmail.com}
\affiliation{Physics Department, Brookhaven National Laboratory, Upton, New York 11973, USA}

\author{Huey-Wen Lin}
\email{hwlin@pa.msu.edu}
\affiliation{Department of Physics and Astronomy, Michigan State University, East Lansing, MI 48824}
\affiliation{Department of Computational Mathematics,
  Science and Engineering, Michigan State University, East Lansing, MI 48824}

\preprint{MIT-CTP/4876}
\preprint{MSUHEP-17-003}




\begin{abstract}
We present the first lattice-QCD calculation of the pion distribution amplitude using the large-momentum
effective field theory (LaMET) approach, which allows us to extract lightcone parton observables from a Euclidean lattice. 
The mass corrections needed to extract the pion distribution amplitude from this approach are calculated to all orders in $m^2_{\pi}/P_z^2$.
We also implement the Wilson-line renormalization which is crucial to remove the power divergences in this approach,  
and find that it reduces the oscillation at the end points of the distribution amplitude. Our exploratory result at 310-MeV pion mass favors a single-hump form broader than the asymptotic form of the pion distribution amplitude. 
\end{abstract}

\maketitle

\section{Introduction}\label{sec:intro}

Hadronic lightcone distribution amplitudes (DAs) play an essential role in the description of hard exclusive processes involving large momentum transfer. They are crucial inputs for processes relevant to measuring fundamental parameters of the Standard Model and probing new physics~\cite{Stewart:2003gt}. The QCD factorization theorem and asymptotic freedom allow us to separate the short-distance physics incorporated in the hard quark and gluon subprocesses from the long-distance physics incorporated in the process-independent hadronic DAs. 
While the short-distance hard quark and gluon subprocesses are calculable perturbatively, the hadronic DAs are intrinsically nonperturbative. To determine them, 
we must resort to experimental measurements, lattice calculations or QCD models. 

The simplest and most extensively studied hadronic DA is the twist-2 DA of the pion. It represents the probability amplitude of finding the valence $q\bar q$ Fock state in the pion with the quark (antiquark) carrying a fraction $x$ ($1-x$) of the total pion momentum. 
The pion lightcone distribution amplitude (LCDA) is defined as 
\beq\label{LCDA}
\phi_{\pi}(x)={\frac{i}{f_{\pi}}}\int\frac{d\xi}{2\pi}e^{i(x-1)\xi\lambda\cdot P}\langle \pi(P)|\bar \psi(0)\lambda\cdot\gamma\gamma_5 \Gamma(0,\xi\lambda)\psi(\xi\lambda)|0\rangle
\eeq
with the normalization $\int_0^1 dx\, \phi_\pi(x)=1$, where the two quark fields are separated along the lightcone with $\lambda^\mu=(1,0,0,-1)/\sqrt 2$, and $x$ ($1-x$) denotes the momentum fraction of the quark (antiquark).
The twist-2 pion DA can be constrained from experimental measurements of e.g. the pion form factor~\cite{Farrar:1979aw}%
, and then as an input can be used to test QCD in, for example, $\gamma \gamma* \rightarrow \pi^0$ from BaBar and Belle~\cite{Aubert:2009mc,Uehara:2012ag}.
Some experiments proposed~\cite{Sawada:2016mao} at J-PARC might also be of use.
At large momentum transfer, the pion DA is well known to follow a universal asymptotic form~\cite{Lepage:1979zb}:  
$\phi_\pi(x, \mu \to \infty) \rightarrow 6 x(1-x)$.
However, 
there have been some debates over the shape of the pion DA at lower scales $\mu$.
For example, Ref.~\cite{Chernyak:1981zz} suggested a ``double-humped'' shape for the pion DA, which is very different from the asymptotic form, while other QCD models (for example, large-$N_c$ Regge model~\cite{RuizArriola:2006jge}, 
QCD sum rule calculations~\cite{Radyushkin:1994xv},
Nambu-Jona-Lasinio model~\cite{RuizArriola:2002bp},
Dyson-Schwinger equations~\cite{Chang:2013pq}, truncated Gegenbauer expansion~\cite{Agaev:2012tm}, just to name a few) 
do not suggest such a feature. Unfortunately, lattice calculations have traditionally only been able to extract the lowest few moments of the pion DA after using the operator product expansion (OPE). The highest moment ever calculated on the lattice is the second moment~\cite{Braun:2015axa,Arthur:2010xf,Braun:2006dg,Daniel:1990ah,Martinelli:1987si}, and most calculations struggled with the noise-to-signal ratio. Ref.~\cite{Cloet:2013tta} took the moment results from lattice-QCD calculations and reconstructed the pion DA using a specific parametrization; however, the errors propagating from the lattice calculations are relatively large, preventing them from discriminating between the QCD models. Calculating moments beyond the lowest two on the lattice is much more difficult due to the breaking of rotational symmetry by discretization, which induces divergent mixing coefficients to lower moments such that the noise-to-signal becomes a big problem. It was proposed to use a smeared source to reduce the discretization error~\cite{Davoudi:2012ya}, or to use another scale to replace the lattice cut-off in the mixing. For example, by using a heavy-light current in the OPE for the current-current correlator, the scale in the mixing parameters is replaced by the heavy-quark mass~\cite{Detmold:2005gg} or the gradient-flow scale in the proposal of Ref.~\cite{Monahan:2016bvm}.
Having an alternative approach to calculate the pion DA with better precision and quantifiable systematics is highly desirable so that it can be used to make predictions in other harder-to-calculate processes, such as $B\rightarrow \pi\pi$.


Recently, a new approach has been proposed to calculate the full 
$x$ dependence of parton quantities, such as parton distributions, distribution amplitudes, etc.~\cite{Ji:2013dva}. The method is based on the observation
that, while in the rest frame of the nucleon, parton physics corresponds to
lightcone correlations, the same physics can be obtained through
time-independent spatial correlations in the infinite-momentum frame (IMF) of the hadron after a matching procedure. For
finite but large momenta feasible in lattice simulations, a large-momentum
effective field theory (LaMET) can be used to relate Euclidean
quasi-observables to physical observables through a factorization theorem~\cite{Ji:2014gla} (there exist also other approaches to extract lightcone quantities from Euclidean ones, see e.g.~\cite{Braun:2007wv,Liu:1993cv,Liu:1998um,Liu:1999ak,Liu:2016djw}). 
Since then, there have been many follow-up studies on factorization~\cite{Ma:2014jla} and determinations of the
one-loop corrections needed
to connect finite-momentum quasi-distributions to lightcone distributions 
for nonsinglet leading-twist PDFs~\cite{Xiong:2013bka}, generalized parton distributions (GPDs)~\cite{Ji:2015qla}, transversity GPDs~\cite{Xiong:2015nua} and pion DA~\cite{Ji:2015qla} in the continuum. Reference~\cite{Ji:2015jwa} also explores the renormalization of quasi-distributions, and establishes that the quasi-distribution is multiplicatively renormalizable at two-loop order. 
There are also proposals to improve the quark correlators to remove linear divergences in the one-loop matching~\cite{Li:2016amo}, to improve the nucleon source to get higher nucleon momenta on the lattice~\cite{Bali:2016lva}, and to use the non-perturbative evolution of quasi-distributions as a guide for the extrapolation of lattice results at moderate momentum to infinite momentum~\cite{Radyushkin:2016hsy,Radyushkin:2017gjd}. In Refs.~\cite{Ishikawa:2016znu,Chen:2016fxx}, it was shown that the power divergence present in the long-link matrix elements can be removed by a mass renormalization in the auxiliary $z$-field formalism, in the same way as the renormalization of power divergence for an open Wilson line. After the Wilson-line renormalization, the long-link matrix elements are improved such that they contain at most logarithmic divergences. A nonperturbative determination of the mass counterterm can, for example, be done following the procedure based on the static-quark potential for the renormalization of Wilson loop in Ref.~\cite{Musch:2010ka}. 

The first attempts to apply the LaMET approach to compute parton observables were the direct lattice computations of the unpolarized, helicity and transversity isovector quark distributions~\cite{Lin:2014gaa,Lin:2014yra,Lin:2014zya,Alexandrou:2015rja,Chen:2016utp,Alexandrou:2016jqi}. 
Although the current lattice systematics are not yet fully accounted for, a sea-flavor asymmetry has been qualitatively seen in both the unpolarized and linearly polarized cases, part of which 
has been confirmed in the updated measurements by the STAR~\cite{Adamczyk:2014xyw} and PHENIX~\cite{Adare:2015gsd}
collaborations. The Drell-Yan experiments at FNAL (E1027+E1039) and future EIC data will be able to give more insight into the sea asymmetry in the transversely polarized nucleon. 

In this paper, we present the first direct lattice-QCD results for the Bjorken-$x$ dependence  of the pion DA 
using lattice gauge ensembles with $N_f=2+1+1$ highly improved staggered
quarks (HISQ)~\cite{Follana:2006rc} (generated by the MILC Collaboration~\cite{Bazavov:2012xda}) and clover valence fermions with pion mass $310$~MeV.
In the framework of LaMET, the pion LCDA $\phi(x)$ can be studied from the IMF limit of the following quasi-correlation 
\beq\label{quasiDA}
{\tilde \phi}(x, P_z)={\frac{i}{f_{\pi}}}\int\frac{dz}{2\pi}e^{-i(x-1)P_z z}\langle \pi(P)|\bar\psi(0)\gamma^z\gamma_5 \Gamma(0,z)\psi(z)|0\rangle
\eeq
with the two quark fields separated along the spatial $z$ direction. As shown in Ref.~\cite{Ji:2015qla}, the pion LCDA can be related to the quasi-DA by the following matching formula
\beq\label{pionDA1loopmatching}
{\tilde \phi}(x, \Lambda, P_z)=\int_0^1 dy\, Z_\phi(x, y, \Lambda, \mu, P_z)\phi(y, \mu)+\mathcal{O}\left(\frac{\Lambda^2_{QCD}}{P_z^2},\frac{m^2_{\pi}}{P_z^2}\right) ,
\eeq
where $\Lambda=\pi/a$ is the UV cutoff for the quasi-DA with $a$ the lattice spacing. $\mu$ denotes the $\overline{\text{MS}}$ renormalization scale of the pion LCDA. Using Eq.~\ref{pionDA1loopmatching}, we will be able to recover the pion LCDA.

The paper is organized as follows: We will 
start by discussing the finite-momentum corrections for the quasi-DA computed on the lattice in Sec.~\ref{sec:corrections}, and then present the lattice results in Sec.~\ref{sec:num}. We first show the results without Wilson-line renormalization to remove the power divergence, and then explore the impact of Wilson-line renormalization where the mass counterterm is determined by using the static-quark potential for the renormalization of Wilson loop discussed in Ref.~\cite{Musch:2010ka}. Finally we summarize in Sec.~\ref{sec:sum}. The details of the finite-momentum corrections are given in the Appendices.

\section{Finite-$P_z$ Corrections for Pion Distribution Amplitude}\label{sec:corrections}

In this section, we present the finite-momentum corrections needed for the calculation of pion DA. In the limit $P_z \rightarrow \infty$, the matching becomes the most
important $P_z$ correction. The factor
$Z_\phi$
has been computed up to one loop in
Ref.~\cite{Ji:2015qla} using a momentum-cutoff regulator instead of a
lattice regulator. Therefore, this $Z$ factor is accurate up to the
leading logarithm but not for the numerical constant. Determining this
constant requires a calculation using lattice perturbation theory with the same lattice
action.

At tree level, the $Z_\phi$ factor is just a delta function. Up to one-loop level, we can write
\begin{equation}
Z_\phi(x, y) = \delta (x-y) + \frac{\alpha_s}{2\pi} \overline{Z}_\phi(x, y)
  + \mathcal{O}\left(\alpha_s^2 \right),
\end{equation}
such that
\begin{equation}
\tilde{\phi}(x) \simeq \phi(x)
  + \frac{\alpha_s}{2\pi} \int dy\,
    \overline{Z}_\phi\!\left(x, y\right)\phi(y).
\end{equation}
Since the difference between $\tilde{\phi}(x)$ and $\phi(x)$ starts at
the loop level, we can rewrite the above equation as
\begin{equation}
\phi(x) \simeq \tilde{\phi}(x)
  - \frac{\alpha_s}{2\pi} \int dy\,
    \overline{Z}_\phi\!\left(x, y\right)\tilde{\phi}(y)
\label{q}
\end{equation}
with an error of $\mathcal{O}\left(\alpha_s^2\right)$~\cite{Ma:2014jla}. As in the parton distribution, $\overline{Z}_\phi(x, y)$ can be written as
\begin{equation}\label{eq7}
\overline{Z}_\phi(x, y) = \left(Z_\phi^{(1)}(x, y) - C\delta(x-y)\right),
\end{equation}
with the first term coming from gluon emission and the second term from the quark
self-energy diagram, $C=\int_{-\infty}^{\infty} d x'\,Z_\phi^{(1)}(x', y)$. (This implies $\int dx \phi(x)=\int dx \tilde{\phi}(x)$ at one loop, which follows from the conservation of the non-singlet axial current when quark masses are neglected.) Using this, Eq.~\ref{q} becomes
\begin{equation}
\phi(x) \simeq \tilde{\phi}(x)
  - \frac{\alpha_s}{2\pi} \int_{-\infty}^{\infty}\!dy\,
  \left[ Z_\phi^{(1)}\!\left(x, y\right)
    \tilde{\phi}(y)
  - Z_\phi^{(1)}\!\left(y, x\right)
    \tilde{\phi}(x)\right] ,  \label{cov}
\end{equation}
where for simplicity we have extended the integration range of $y$ to infinity, which introduces an error at higher order. The expression for the matching factor $Z_\phi^{(1)}(x, y)$ is given in Appendix~A.

For a finite $P_z$, we need to take into account {the $\mathcal{O}\left(m^2_{\pi}/P_z^2\right)$
meson-mass and $\mathcal{O}\left(\Lambda^2_\text{QCD}/P_z^2\right)$} higher-twist corrections. Following a procedure similar to Ref.~\cite{Chen:2016utp}, we can derive the mass corrections to all orders in $m^2_{\pi}/P_z^2$, which leads to the following relation between the pion DAs (for details see Appendix~B).
\begin{align}\label{masscorr}
\phi(x)&=\sqrt{1+4c}\sum_{n=0}^\infty \frac{(4c)^n}{f_+^{2n+1}}\Big[(1+(-1)^n)\tilde\phi\Big(\frac{1}{2}-\frac{f_+^{2n+1}(1-2x)}{4(4c)^n}\Big)+(1-(-1)^n)\tilde\phi\Big(\frac{1}{2}+\frac{f_+^{2n+1}(1-2x)}{4(4c)^n}\Big)\Big],
\end{align}
where $c=m_\pi^2/4P_z^2$ and $ f_{+}=\sqrt{1+4c}+ 1$.

The $\mathcal{O}\left(\Lambda^2_\text{QCD}/P_z^2\right)$ correction can be derived in the same way as in Ref.~\cite{Chen:2016utp}, since the twist-4 operator involved is the same. The twist-4 effect can be implemented by adding a
$\tilde{\phi}_\text{twist-4}$ contribution to $\tilde{\phi}$, such that
\begin{equation}\label{highertwist}
\tilde{\phi}(x,\Lambda,P_z) \rightarrow
  \tilde{\phi}(x,\Lambda,P_z) + \tilde{\phi}_\text{twist-4}(x,\Lambda,P_z),
\end{equation}
where
\begin{equation}
\tilde{\phi}_\text{twist-4}(x,\Lambda,P_z) =
  \frac{1}{8\pi}\int_{-\infty}^\infty\!dz\,\Gamma_0 \left(-ixzP_z\right)
  \left\langle \pi(P)\left\vert \mathcal{O}_\text{tr}(z)\right\vert 0\right\rangle,
  \label{t4-1}
\end{equation}
$\Gamma_0$ is the incomplete Gamma function and
\begin{align}
\mathcal{O}_\text{tr}(z) &=
  \int_0^z\!dz_1\,\bar{\psi}(0) \Big[ \gamma^\nu\gamma_5 \Gamma\left(0,z_1\right)
    D_\nu\Gamma \left(z_1,z\right) \notag \\
&{} + \int_0^{z_1}\!dz_2\, \lambda \cdot \gamma\gamma_5
  \Gamma\left(0,z_2 \right) D^\nu \Gamma \left(z_2,z_1\right)
  D_\nu \Gamma\left(z_1,z\right) \Big] \psi (z\lambda)  \label{t4-2}
\end{align}
with $\lambda^\mu=(0,0,0,-1)$. Eqs.~\ref{cov}--\ref{highertwist} take into account the one-loop, mass and higher-twist corrections, respectively. We need to implement them step by step to achieve the final pion DA. For the higher-twist corrections, instead of computing them directly on the lattice, we only
parametrize and fit them as a $1/P_z^2$ correction after we have
removed other leading-$P_z$ corrections, as was done in Ref.~\cite{Chen:2016utp}.

%

\section{Numerical Results and Discussion}\label{sec:num}

In this section, we report the first results of a lattice-QCD calculation of the $x$-dependence of the pion DA.
We use clover valence fermions on gauge ensembles with $2+1+1$ flavors (degenerate up/down, strange and charm degrees of freedom in the QCD vacuum) of highly improved staggered quarks (HISQ)~\cite{Follana:2006rc} generated by MILC Collaboration~\cite{Bazavov:2012xda}. The pion 
mass of this ensemble is $m_\pi \approx 310$~MeV with lattice spacing $a\approx 0.12$~fm and box size $L\approx 3$~fm, corresponding to $m_\pi L \approx 4.5$.
The HISQ ensembles are hypercubic (HYP)-smeared~\cite{Hasenfratz:2001hp} and the clover parameters are tuned to recover the lowest pion mass of the staggered quarks in the sea.\footnote{Other studies using the same setup are done in Refs.~\cite{Bhattacharya:2016zcn,Bhattacharya:2015wna,Bhattacharya:2015esa,Bhattacharya:2013ehc} and no exceptional-configuration behavior was observed.}   
HYP smearing has been shown to significantly improve the discretization effects on operators and shift their corresponding renormalizations toward their tree-level values (near 1 for quark bilinear operators). 
The results shown in this work are done using correlators calculated from 3 source locations on 986 configurations.  For each positive $z$-momentum $P_z$, the matrix elements are averaged with their corresponding $-P_z$ to improve the signal. 

\subsection{Pion Quasi-Distribution Amplitude}

\begin{figure}[tbp]
\includegraphics[width=0.6\textwidth]{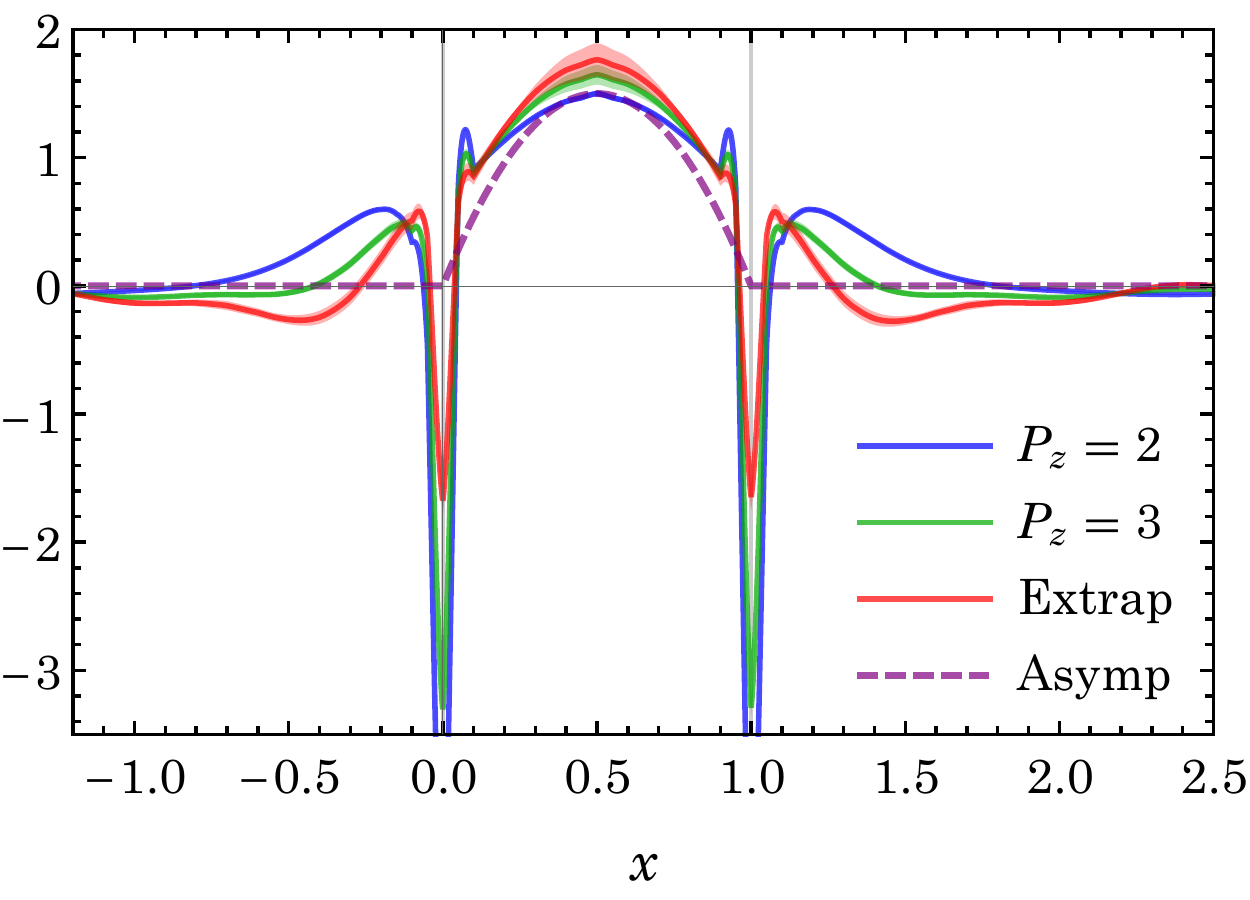}
\caption{The pion quasi-distribution amplitude (at $\mu=2$~GeV) after one-loop and mass correction for $P_z=2$ (blue) and $3$ (green) (in units of $2\pi/L$).
The extrapolation to infinite momentum to remove the remaining higher-twist effects is shown in red. The Wilson-line renormalization that removes the power divergent contribution is not included in this plot, and will be implemented later in the results of improved pion quasi-DA. The purple dashed line is the asymptotic form $6x(1-x)$.
}
\label{fig:pionDA}
\end{figure}

We begin with the pion quasi-DA without the Wilson-line renormalization. Here, we follow similar steps to those listed in our previous work on nucleon parton distribution functions: First, we implement the one-loop and mass corrections whose formulae are detailed in the previous sections, and extrapolate to the infinite-momentum limit via {$\alpha(x)+\beta(x)/P_z^2$} (and thereby remove the higher-twist terms that come in at $O(\Lambda_\text{QCD}^2/P_z^2)$). The true light-cone pion DA should be recovered. 
Fig.~\ref{fig:pionDA} shows the results for the 
pion quasi-DA at $\mu=2$~GeV after including one-loop and mass corrections at different momenta $P_z=2, 3$ (in units of $2\pi/L$)\footnote{For this work, we initially calculate the pion quasi-DA for 3 momenta, $P_z= 1, 2, 3$ (in units of $2\pi/L$), but the corrections term for the smallest-momentum distribution is less well-behaved, as observed in the nucleon PDF case~\cite{Chen:2016utp}; thus, we drop it in the rest of this work.}. We then extrapolate using these 2 momenta to the infinite-momentum limit using the form {$\alpha(x)+\beta(x)/P_z^2$}, shown in red, where a linear divergence is present in the one-loop matching kernel (later, we will show improved results for the pion DA where the power divergence is removed by taking into account the Wilson-line normalization). The dashed line is the asymptotic form $6x(1-x)$. All our resulting curves are symmetric around $x=1/2$, as expected from the symmetry of the pion DA under the interchange $x\leftrightarrow 1-x$. The pion DA has often been expanded in terms of Gegenbauer polynomials in past studies, and the dashed curve here contains only the zeroth Gegenbauer polynomial. The other three curves are broader than the asymptotic form, indicating contribution from higher Gegenbauer polynomials.

We note several interesting features of this result.
First, the pion DA is expected to vanish outside the region $x\in [0,1]$ after taking the IMF limit. We see the $P_z=2$ pion quasi-DA is nonzero for $x\in [1,1.7]$, and this range shrinks to $x\in [1,1.4]$ for $P_z=3$. A similar pattern is observed for the region $x<0$. The distributions are moving in the right direction as the pion DA will vanish outside $[0,1]$ with $P_z \to \infty$. 
However, after taking the IMF limit via extrapolation formula $\alpha(x)+\beta(x)/P_z^2$, we find there is still residual distribution outside $x\in [0,1]$. 
This is likely due to using the approximation Eq.~\ref{cov}, where the cancellation among $\tilde\phi(x)$ outside the $x\in [0,1]$ region is between an all-order result and a perturbative expression, and is therefore incomplete\footnote{Although the difference here is formally of higher order, it might have a sizable numerical effect.}.
This can be improved by including the higher-order matching and going to larger momentum, which we will explore more extensively in future work.

Second, the results near $x=0$ and $x=1$ are not reliable. There are unphysical peaks and dips due to the linear divergence in the one-loop matching in these regions, which become smaller as $P_z$ becomes larger. The smallest-$x$ region is dominated by the smallest nonzero momentum fraction, which is proportional to $1/L$ (where $L$ is the lattice length along boosted-momentum direction), due to the finite box size. To improve results near these regions would require large momentum and large box size.

Third, the unphysical oscillatory behavior near $x=0$ and $x=1$ is largely due to the presence of a linear divergence in the one-loop matching for the bare long-link matrix element. In Refs.~\cite{Ishikawa:2016znu,Chen:2016fxx}, it has been shown that the power divergence (in the $a \to 0$ limit) in the long-link operator can be removed to all orders by a mass counterterm $\delta m$ (in the auxiliary $z$-field description of the Wilson line), which is the same as in the renormalization of an open Wilson line. After
the Wilson-line renormalization, the pion quasi-DA is improved such that it contains at most logarithmic
divergences. We will investigate this improved quasi-DA numerically in the rest of the paper. 

\subsection{Improved Pion Quasi-Distribution Amplitude}
The improved pion quasi-DA without power divergence can be defined as~\cite{Chen:2016fxx}
\beq\label{impDA}
{\tilde \phi}_{\text{imp}}(x, P_z)=\frac{i}{f_\pi}\int\frac{dz}{2\pi}e^{-i(x-1)P_z z-\delta m |z|}\langle \pi(P)|\bar\psi(0)\gamma^z\gamma_5 \Gamma(0,z)\psi(z)|0\rangle ,
\eeq
where $\delta m$ should be determined nonperturbatively through studying the Wilson-line renormalization. It is worthwhile to comment that since the mass counterterm $\delta m$ cancels all power divergence in the pion quasi-DA\footnote{At perturbative one-loop, it appears as a linear divergence, but more-divergent power divergences can appear at higher loops.}, when we do the perturbative matching between Eqs.~\ref{impDA} and \ref{LCDA}, we need to remove the linear divergence present in the one-loop matching kernel for consistency. Moreover, as shown in Ref.~\cite{Chen:2016fxx} and below, $\delta m$ is negative, the exponential factor $e^{-\delta m |z|}$ then increases the weight of matrix elements with relatively large
$z$, and thereby increases the contribution at relatively small momentum when Fourier
transforming to momentum space. It is therefore important to properly account for the higher-twist corrections.

\begin{figure}[tbp]
\includegraphics[width=0.6\textwidth]{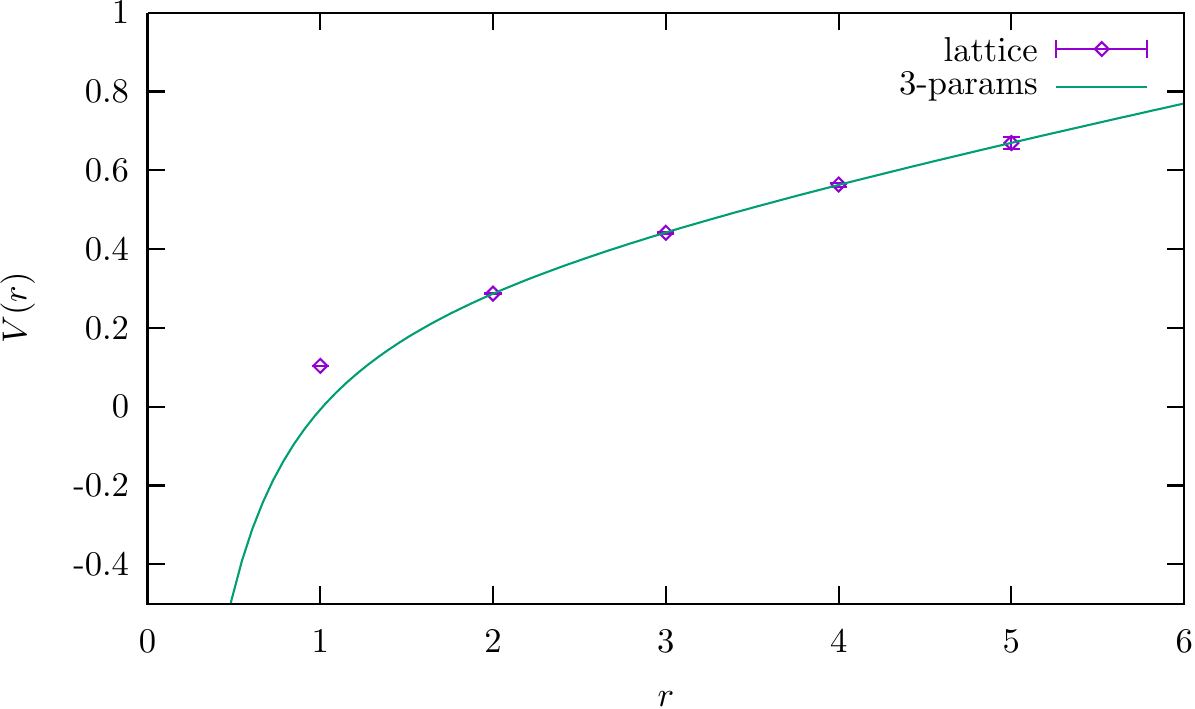}
\caption{The energy of the static-quark pairs fit to the functional form of Eq.~\ref{V}.
The point at $r=1$ is excluded from the fit to reduce discretization error. If we further exclude the $r=2$ point, then $c_2$ is increased by 15\%, still in the range of Eq.~\ref{dm}.}
\label{fig:V}
\end{figure}

We first explore the nonperturbative determination of $\delta m$ discussed in Ref.~\cite{Musch:2010ka} using the static-quark potential for the renormalization of Wilson loop. The Wilson loop $W(t,r)$ of width $r$ and length $t$ is long in the $t$-direction such that higher excitations are sufficiently suppressed. The quark potential is then obtained as
\beq
V(r)=-\frac{1}{a}\lim_{t\to\infty}\ln\frac{\langle \text{Tr}[W(t,r)]\rangle}{\langle \text{Tr}[W(t-a,r)]\rangle},
\eeq
where $a$ is the lattice spacing and the cusp anomalous dimensions from the four sharp corners of the Wilson loop are canceled between numerator and denominator. When $r$ is larger than the confinement scale but shorter than the string breaking scale\footnote{The onset of string breaking can be estimated by $V(r) > 2 m_B-m_{\Upsilon}=1.1$~GeV.}, the lattice data should be described by
the energy of the static quark pairs
\beq\label{V}
V(r)=\frac{c_1}{r}+c_2+c_3 r ,
\eeq
where the $c_1$ term is the Coulomb potential which dominates at short distance,  $c_3$ term is the confinement linear potential. The $c_2$ term is twice the rest mass of the heavy quark, and we expect $c_2=\tilde{c}/a+\mathcal{O}(\Lambda_{\text{QCD}})$. Thus, the $\delta m$ counterterm that cancels the linear divergence in the Wilson line is 
\beq
\delta m=-\frac{\tilde{c}}{2a}= -\frac{c_2}{2} +\mathcal{O}(\Lambda_{\text{QCD}}).
\eeq
This leads to
\beq\label{dm}
\delta m\simeq -260 \pm 200 \mbox{ MeV},
\eeq
where we have used the fitted value $\delta m=-0.16/a$ from Fig.~\ref{fig:V}, which is $0.38$ times of the one-loop value computed in Ref.~\cite{Chen:2016fxx}, 
and we estimate the error by the size of $\Lambda_{\text{QCD}} \sim 200$~MeV. The error can be reduced by performing the computation at different $a$ to extract the $1/a$-dependent term in $c_2$.

As mentioned before, once the improved pion DA of Eq.~\ref{impDA} is used with $\delta m$ determined nonperturbatively, the linear divergence in the one-loop matching kernel will {be canceled} by the $\delta m$ counterterm as shown in Eq.~\ref{imprmat}.
In Ref.~\cite{Chen:2016fxx}, it was demonstrated that in the limit $\Lambda/P_z \to \infty$, only the Wilson-line self-energy diagram is divergent among the ``real diagrams'' (i.e. $Z_\phi^{(1)}(x, y)$ of Eq.~\ref{eq7}) 
in one loop and in the Feynman gauge. Therefore, in a lattice perturbation theory calculation, one only needs to calculate this diagram, which is linearly divergent ($\propto \Lambda/P_z$). Using the simplest version of gauge-field discretization, one finds the matching between the momentum and lattice cut-offs is $\Lambda =\pi/a +\mathcal{O}(a^2)$. This result holds not only for the non-singlet quasi-PDF operator used in Ref.~\cite{Chen:2016fxx}, but also for the pion quasi-DA in this work. The ``virtual diagrams'' (i.e. $C$ of Eq.~\ref{eq7}) will contain logarithmic divergence from the quark self-energy diagram, which can be removed by adding counterterms in the lattice action or treating the integration limits of $C$ carefully. In Eq.~\ref{imprmat}, the $\Lambda/P_z \to \infty$ limit is not taken, so $C$ is finite. We find that the difference between taking this limit and not is small, certainly within the error induced by the uncertainty of $\delta m$.

\begin{figure}[tbp]
\includegraphics[width=0.6\textwidth]{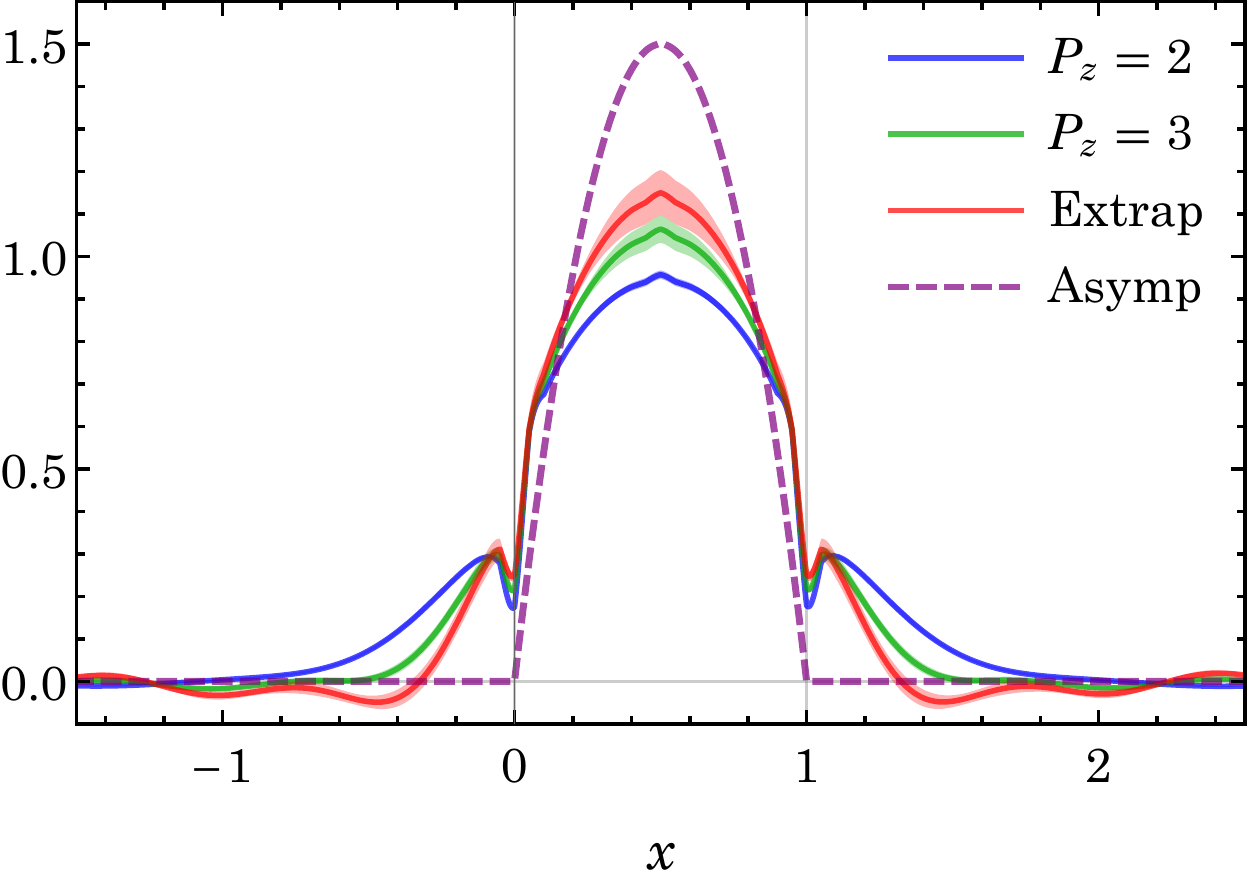}
\caption{The improved pion distribution amplitude at $\mu=2$~GeV using $\delta m=0.38\delta m_{\text{1-loop}}$ in Eq.~\ref{impDA} for $P_z=2$ (blue) and $3$ (green) (in units of $2\pi/L$) 
and extrapolation to infinite-momentum limit (red), along with the asymptotic form $6x(1-x)$ (dashed line).
}
\label{fig:pionDApheno}
\end{figure}

The resulting improved pion quasi-DA using Eq.~\ref{impDA} 
and the central value of $\delta m$ is shown in Fig.~\ref{fig:pionDApheno}. The unphysical oscillations near $x=0$ and $x=1$ are largely removed. There are still small kinks in the unphysical region, but they are expected to vanish when higher-order matching is taken into account and the $P_z \to \infty$ limit is approached. 

\begin{figure}[tbp]
\includegraphics[width=0.65\textwidth]{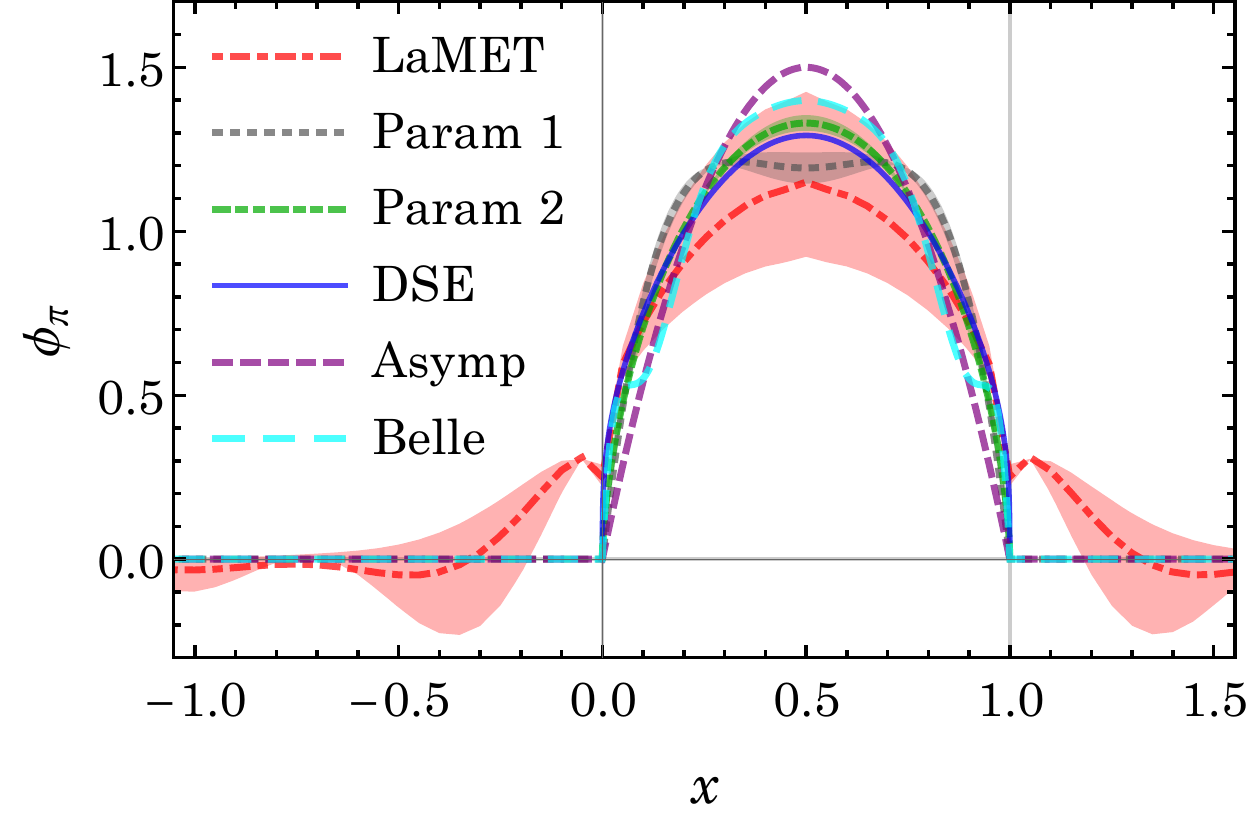}
\caption{The improved pion distribution amplitude at $\mu=2$~GeV with $\delta m=(0.38 \pm 0.28)\delta m_{\text{1-loop}}$ (red band with the central value denoted by red dot-dashed) obtained in this work (labeled as ``LaMET''), along with that obtained from the Dyson-Schwinger equation (labeled ``DSE'') analysis of the pion (blue), a fit to the Belle data (labeled ``Belle", cyan), parametrized fits to the lattice moments (labeled ``Param 1" and ``Param 2'', respectively, gray and green) and the asymptotic form (labeled ``Asymp", purple).
}
\label{fig:x}
\end{figure}

The final result that includes the lattice statistical uncertainties, finite-$P_z$ corrections and the uncertainty of $\delta m$ estimated in Eq.~\ref{dm} is presented in Fig.~\ref{fig:x}. 
Also shown in the same figure are the model calculation from the Dyson-Schwinger equation (DSE)~\cite{Chang:2013pq}, from the truncated Gegenbauer expansion fit to the Belle data for the $\gamma\gamma^*\to\pi^0$ form factor~\cite{Agaev:2012tm} and from parametrizations of the pion DA with the parameters fit to lowest-moment calculations from lattice QCD in~\cite{Braun:2015axa}. For the fit to the Belle data, we use the Gegenbauer polynomial expansion up to the eighth moment given in Ref.~\cite{Agaev:2012tm} and run to 2~GeV. For the fit to the lattice moment calculations, we have chosen two different parametrizations. One is simply a truncation of the Gegenbauer polynomial expansion of the pion DA to the second order $\phi(x)=6x(1-x)[1+a_2 C_2^{3/2}(2x-1)]$ (labeled ``Param 1") with the value of $a_2$ taken from~\cite{Braun:2015axa}. The other is $\phi(x)=A[x(1-x)]^B$ with $A$ and $B$ determined from the normalization condition and the second moment of the pion DA (labeled ``Param 2"). The second parametrization is close to the DSE result, but differs from the first parametrization. The difference between them can be viewed as a rough estimate of errors from the truncation, and reflects uncertainties in the parametrization, which are currently underestimated even though both bands have smaller errors than ours. A direct calculation of the $x$-dependence will help to resolve such uncertainties. Of course, this can be achieved only when the direct calculation reaches a sufficiently high accuracy, which is difficult at the current stage but might be improved in the foreseeable future. Nonetheless,
the results of our direct calculation at 310-MeV pion mass is in agreement within errors with DSE, Belle data fit result and the parametrized reconstruction of pion DAs in the region near $x=1/2$, although the two parametrized forms differ from each other.
The uncertainty of our distribution is dominated by the $\delta m$ uncertainty, which can be largely removed by performing calculations at different lattice spacing. 
As before, we still have residual distribution outside the $[0,1]$ region, which should vanish when larger momenta are reached and higher-order matching is taken into account in the future. 
Also, as is typical in an exploratory study, the pion mass in this work is still heavier than its physical value. However, the study of Ref.~\cite{Chen:2003fp} shows that the leading chiral correction for $\phi_{\pi}(x)$ is proportional to $m_{\pi}^2$ with the chiral logarithm $m_{\pi}^2 \ln m_{\pi}^2$ completely absorbed by $f_{\pi}$. This property will simplify the chiral extrapolation in future computations.
It is encouraging that our current result is qualitatively similar to other determinations using lattice-moment parametrization, models and fits to experimental data, and also favors a single-hump distribution in $\phi_{\pi}(x)$. 

\section{Summary and Outlook}\label{sec:sum}
In this work, we presented the first lattice-QCD calculation of the pion distribution amplitude using the large-momentum
effective field theory (LaMET) approach. 
We derived the mass-correction formulation needed for the pion quasi-distribution amplitude. 
We also implemented the Wilson-line renormalization in this work, which is important to remove the power divergences in LaMET approach; and found that it reduces the oscillation at the end points of the distribution amplitude.
Finally, our result at 310-MeV pion mass shows similar behavior as previous studies done using DSE, a fit to the Belle data and as parametrizations with latest lattice moment result, and favors a single-hump structure.

However, in the current study, we have not accounted for all possible systematic uncertainties, and there are multiple improvements that can be done in future studies. For example, in our work, it is clear that larger boosted momentum is needed for the pion distribution amplitude to make the result outside the physical region consistent with $0$ than for the unpolarized nucleon parton distribution function. Finer lattice spacing would help reduce the uncertainty in the counterterm determined by the Wilson-loop study. Larger lattice box and also higher-order matching would reduce the unphysical kinks near $x=1$ and $0$. Last but not least, we hope this work will encourage following works to extensively study the distribution amplitude of the pion and other hadrons. 

\section*{Acknowledgments}

JHZ thanks G. Bali, V. Braun and M. G\"ockeler and A. Sch\"afer for valuable discussions and comments. The LQCD calculations were performed using the Chroma software
suite~\cite{Edwards:2004sx}. We thank MILC Collaboration for sharing the lattices used to perform this study. 
Computations for this work were carried out in part on facilities of the
USQCD Collaboration, which are funded by the Office of Science of the
U.S. Department of Energy, and on the National Energy Research Scientific Computing Center, a DOE Office of Science User Facility supported by the Office of Science of the U.S. Department of Energy under Contract No. DE-AC02-05CH11231.
This work was partially supported by the U.S. Department of Energy via grants DE-FG02-93ER-40762, a grant (No.~11DZ2260700) from the Office of Science and Technology in Shanghai Municipal Government, grants from National Science Foundation of China (No.~11175114, No.~11405104, No.~11655002), a DFG grant SCHA~458/20-1, the SFB/TRR-55 grant "Hadron Physics from Lattice QCD", the MIT MISTI program, the Ministry of Science and Technology, Taiwan under Grant Nos.~105-2112-M-002-017-MY3 and 105-2918-I-002 -003 and the CASTS of NTU. The work of JWC and XJ is supported in part by the U.S. Department of Energy, Office of Science, Office of Nuclear Physics, within the framework of the TMD Topical Collaboration.

\section*{Appendix A: One-Loop Matching for Quasi-DA of Pion}\label{sec:oneloop_appendix}

In this Appendix, we list the one-loop matching factors used throughout this paper. These factors have been obtained in Ref.~\cite{Ji:2015qla}. However, as in Ref.~\cite{Chen:2016utp}, we keep a finite cutoff $\Lambda$ and do not take the limit $\Lambda\gg x P_z$.

For the pion distribution amplitude, expanding the matching factor $Z_\phi(x, y, \Lambda, \mu, P_z)$ in Eq.~(\ref{pionDA1loopmatching}) as
\beq
Z_\phi(x, y, \Lambda, \mu, P_z)=\delta(x-y)+\frac{\alpha_S}{2\pi}Z_\phi^{(1)}(x, y, \Lambda, \mu, P_z)+\dots,
\eeq
we have
\begin{align}\label{DAmatfac}
Z_\phi^{(1)}(x, y, \Lambda, \mu, P_z)/C_F&=G_1(x, y, \Lambda, \mu, P_z)\theta(x<0)+G_2(x, y, \Lambda, \mu, P_z)\theta(0<x<y)\non\\
&+G_3(x, y, \Lambda, \mu, P_z)\theta(y<x<1)+G_4(x, y, \Lambda, \mu, P_z)\theta(x>1)
\end{align}
with
\begin{align}\label{DAmatfac1}
G_1(x, y, \Lambda, \mu, P_z)&=\frac{1}{x-y}+\frac{\Lambda(x,1)-\Lambda(x,y)}{2P_z(1-y)(x-y)}+\frac{\Lambda(x,y)-\Lambda(x,0)}{2P_z (x-y)y}+\frac{\Lambda(x,y)+\Lambda(y,x)}{2P_z(x-y)^2}\non\\
&+\big(\frac{1}{x-y}-\frac{x}{1-y}\big)\ln(1-x)+\big(\frac{1}{x-y}+\frac{1-x}{y}\big)\ln(-x)+\big(\frac{x}{1-y}-\frac{1-x}{y}-\frac{2}{x-y}\big)\ln(y-x)\non\\
&-\big(\frac{1-x}{2y}+\frac{1}{2(x-y)}\big)\ln\frac{\Lambda(0,x)}{\Lambda(x,0)}+\big(\frac{x}{2(1-y)}-\frac{1}{2(x-y)}\big)\ln\frac{\Lambda(1,x)}{\Lambda(x,1)}\non\\
&+\big(\frac{x}{2(1-y)}-\frac{1-x}{2y}-\frac{1}{x-y}\big)\ln\frac{\Lambda(x,y)}{\Lambda(y,x)},\non\\
G_2(x, y, \Lambda, \mu, P_z)&=\frac{3}{2y}+\frac{1}{2(y-1)}-\frac{2}{y-x}+\frac{\Lambda(x,0)}{2P_z(y-x)y}+\frac{\Lambda(x,1)}{2P_z(1-y)(x-y)}+\frac{(x+y-2xy)(\Lambda(x,y)+\Lambda(y,x))}{4P_z(x-y)^2 y(1-y)}\non\\
&+\big(\frac{x-1}{y}+\frac{1}{y-x}\big)\ln\frac{P_z^2}{\mu^2}+\big(\frac{x-1}{y}+\frac{1}{y-x}\big)\ln(4x)
+\big(\frac{x}{y-1}-\frac{1}{y-x}\big)\ln(1-x)\non\\
&+\big(\frac{x-1}{y}+\frac{2}{y-x}-\frac{x}{y-1}\big)\ln(y-x)-\big(\frac{1}{2y}+\frac{1}{2(x-y)}\big)\ln\frac{\Lambda(0,x)}{\Lambda(x,0)}\non\\
&+\big(\frac{x}{2(1-y)}-\frac{1}{2(x-y)}\big)\ln\frac{\Lambda(1,x)}{\Lambda(x,1)}+\frac{x}{y}\ln\frac{\Lambda(x,y)}{\Lambda(x,0)}+\big(\frac{x}{2(1-y)}-\frac{1}{2y}-\frac{1}{x-y}\big)\ln\frac{\Lambda(x,y)}{\Lambda(y,x)},\non\\
G_3(x, y, \Lambda, \mu, P_z)&=G_2(1-x, 1-y, \Lambda, \mu, P_z),\non\\
G_4(x, y, \Lambda, \mu, P_z)&=G_1(1-x, 1-y, \Lambda, \mu, P_z),
\end{align}
where $\Lambda(x, y)=\sqrt{\Lambda^2+(x-y)^2 P_z^2}+ (x-y)P_z$.

Near $x=y$, one has an extra contribution from the quark wavefunction renormalization
\beq
Z_\phi^{(1)}(x, y, \Lambda, \mu, P_z)/C_F=\delta Z_\phi^{(1)}(2\pi/\alpha_S)\delta(x-y),
\eeq
where $\delta Z_\phi^{(1)}$ provides a plus prescription for the factor in Eq.~\ref{DAmatfac}, and can be written as
\beq
\delta Z_\phi^{(1)}=\int dx\, Z_\phi^{(1)}(x, y, \Lambda, \mu, P_z).
\eeq
If one used the improved pion DA of Eq.~\ref{impDA} for the computation, then the $G_i$ function in the matching kernel will be replaced by
\beq\label{imprmat}
G_i(x, y, \Lambda, \mu, P_z)\to G_i(x, y, \Lambda, \mu, P_z)- \frac{\Lambda}{P_z(x-y)^2}.
\eeq

\section*{Appendix B: Meson Mass Correction for Quasi-DA of Pion}
\label{sec:mmc_appendix}


In this Appendix, we derive the meson-mass corrections to the quasi-DA of the pion. 
For the pion DA, we need to calculate the same series sum as for the unpolarized parton distribution in Ref.~\cite{Chen:2016utp}:
\beq\label{Kn}
 K_n=\frac{\langle (1-2x)^{n-1} \rangle_{\tilde\phi}}{ \langle (1-2x)^{n-1} \rangle_{\phi}}=\sum_{i=0}^{i_\text{max}} C_{n-i}^i c^i =\frac{\lambda_{(\mu_1}\cdots \lambda_{\mu_n)}
  P^{\mu_1}\cdots P^{\mu_n}}{\lambda_{\mu_1}\cdots \lambda_{\mu_n}
  P^{\mu_1}\cdots P^{\mu_n}},
\eeq
where $c=m_{\pi}^2/4 P_z^2$ and $(\ldots)$ means the indices enclosed are symmetric and traceless. The result for even $n$ ($=2k$) is
\begin{equation}\label{evensum}
\sum_{j=0}^k  C_{n-j}^j c^j =\frac{1}{\sqrt{1+4c}}\Big[\left(\frac{f_-}{2}\right)^{2k+1}+\left(\frac{f_+}{2}\right)^{2k+1}\Big],
\end{equation}
while for odd $n$ ($=2k+1$), it is
\begin{equation}\label{oddsum}
\sum_{j=0}^k C_{n-j}^j c^j =\frac{1}{\sqrt{1+4c}}\Big[-\left(\frac{f_-}{2}\right)^{2k+2}+\left(\frac{f_+}{2}\right)^{2k+2}\Big],
\end{equation}
where $f_{\pm}=\sqrt{1+4c}\pm 1$. 

With Eqs.~\ref{evensum} and \ref{oddsum}, we perform an inverse Mellin transform on the moment relation of Eq.~\ref{Kn}
\begin{equation}
\frac{1}{2\pi i}\int_{-i\infty}^{i\infty} dn \, s^{-n} \langle (1-2x)^{n-1} \rangle .
\end{equation}

To extract $\phi(x)$ from $\tilde\phi(x)$, let us rewrite Eq.~\ref{Kn} for an even $n=2k$ as
\begin{equation}\label{momentinverserel}
\langle (1-2x)^{2k-1}\rangle_\phi=\langle (1-2x)^{2k-1}\rangle_{\tilde \phi}\frac{\sqrt{1+4c}}{\left(\frac{f_-}{2}\right)^{2k+1}+\left(\frac{f_+}{2}\right)^{2k+1}}=\langle (1-2x)^{2k-1}\rangle_{\tilde q}\frac{\sqrt{1+4c}}{\left(\frac{f_+}{2}\right)^{2k+1}}\sum_{n=0}^\infty (-1)^n\left(\frac{f_-}{f_+}\right)^{(2k+1)n}.
\end{equation}
The inverse Mellin transform then leads to
\begin{align}
\phi(x)-\phi(1-x)=2\sqrt{1+4c}\sum_{n=0}^\infty \frac{(-f_-)^n}{f_+^{n+1}}\Big[\tilde\phi\Big(\frac{1}{2}-\frac{f_+^{n+1}(1-2x)}{4f_-^n}\Big)-\tilde\phi\Big(\frac{1}{2}+\frac{f_+^{n+1}(1-2x)}{4f_-^n}\Big)\Big].
\end{align}
Similarly, we have
\begin{equation}
\phi(x)+\phi(1-x)=2\sqrt{1+4c}\sum_{n=0}^\infty \frac{f_-^n}{f_+^{n+1}}\Big[\tilde\phi\Big(\frac{1}{2}-\frac{f_+^{n+1}(1-2x)}{4f_-^n}\Big)+\tilde\phi\Big(\frac{1}{2}+\frac{f_+^{n+1}(1-2x)}{4f_-^n}\Big)\Big].
\end{equation}
Therefore,
\begin{align}
\phi(x)&=\sqrt{1+4c}\sum_{n=0}^\infty \frac{f_-^n}{f_+^{n+1}}\Big[(1+(-1)^n)\tilde\phi\Big(\frac{1}{2}-\frac{f_+^{n+1}(1-2x)}{4f_-^n}\Big)+(1-(-1)^n)\tilde\phi\Big(\frac{1}{2}+\frac{f_+^{n+1}(1-2x)}{4f_-^n}\Big)\Big]\non\\
&=\sqrt{1+4c}\sum_{n=0}^\infty \frac{(4c)^n}{f_+^{2n+1}}\Big[(1+(-1)^n)\tilde\phi\Big(\frac{1}{2}-\frac{f_+^{2n+1}(1-2x)}{4(4c)^n}\Big)+(1-(-1)^n)\tilde\phi\Big(\frac{1}{2}+\frac{f_+^{2n+1}(1-2x)}{4(4c)^n}\Big)\Big],
\end{align}
where in the last line we have used $f_+ f_-=4c$. Since $f_+\gg f_-$ or $c$ and the quasi-DA $\tilde \phi(x)$ vanishes asymptotically for large $x$, the above sum is dominated by the first term with $n=0$. In practical calculations, we can reach reasonable accuracy by taking only the first few terms in the sum. In Refs.~\cite{Braun:2011zr,Braun:2011dg}, it was argued that for hadron-to-vacuum matrix elements, the mass corrections also receive contributions from higher-twist operators that can be reduced to total derivatives of twist-two ones. We do not explicitly consider such terms, since they will anyway be part of the higher-twist corrections that are parametrized with a specific form in the present work.

\ifx\@bibitemShut\undefined\let\@bibitemShut\relax\fi
\makeatother
\bibliography{ref}

\begin{thebibliography}{59}
\expandafter\ifx\csname natexlab\endcsname\relax\def\natexlab#1{#1}\fi
\expandafter\ifx\csname bibnamefont\endcsname\relax
  \def\bibnamefont#1{#1}\fi
\expandafter\ifx\csname bibfnamefont\endcsname\relax
  \def\bibfnamefont#1{#1}\fi
\expandafter\ifx\csname citenamefont\endcsname\relax
  \def\citenamefont#1{#1}\fi
\expandafter\ifx\csname url\endcsname\relax
  \def\url#1{\texttt{#1}}\fi
\expandafter\ifx\csname urlprefix\endcsname\relax\def\urlprefix{URL }\fi
\providecommand{\bibinfo}[2]{#2}
\providecommand{\eprint}[2][]{\url{#2}}

\bibitem[{\citenamefont{Stewart}(2003)}]{Stewart:2003gt}
\bibinfo{author}{\bibfnamefont{I.~W.} \bibnamefont{Stewart}}, in
  \emph{\bibinfo{booktitle}{{Proceedings, 38th Rencontres de Moriond on QCD and
  High-Energy Hadronic Interactions: Les Arcs, France, March 22-29, 2003}}}
  (\bibinfo{year}{2003}), \eprint{hep-ph/0308185}.

\bibitem[{\citenamefont{Farrar and Jackson}(1979)}]{Farrar:1979aw}
\bibinfo{author}{\bibfnamefont{G.~R.} \bibnamefont{Farrar}} \bibnamefont{and}
  \bibinfo{author}{\bibfnamefont{D.~R.} \bibnamefont{Jackson}},
  \bibinfo{journal}{Phys. Rev. Lett.} \textbf{\bibinfo{volume}{43}},
  \bibinfo{pages}{246} (\bibinfo{year}{1979}).

\bibitem[{\citenamefont{Aubert et~al.}(2009)}]{Aubert:2009mc}
\bibinfo{author}{\bibfnamefont{B.}~\bibnamefont{Aubert}} \bibnamefont{et~al.}
  (\bibinfo{collaboration}{BaBar}), \bibinfo{journal}{Phys. Rev.}
  \textbf{\bibinfo{volume}{D80}}, \bibinfo{pages}{052002}
  (\bibinfo{year}{2009}), \eprint{0905.4778}.

\bibitem[{\citenamefont{Uehara et~al.}(2012)}]{Uehara:2012ag}
\bibinfo{author}{\bibfnamefont{S.}~\bibnamefont{Uehara}} \bibnamefont{et~al.}
  (\bibinfo{collaboration}{Belle}), \bibinfo{journal}{Phys. Rev.}
  \textbf{\bibinfo{volume}{D86}}, \bibinfo{pages}{092007}
  (\bibinfo{year}{2012}), \eprint{1205.3249}.

\bibitem[{\citenamefont{Sawada et~al.}(2016)\citenamefont{Sawada, Chang,
  Kumano, Peng, Sawada, and Tanaka}}]{Sawada:2016mao}
\bibinfo{author}{\bibfnamefont{T.}~\bibnamefont{Sawada}},
  \bibinfo{author}{\bibfnamefont{W.-C.} \bibnamefont{Chang}},
  \bibinfo{author}{\bibfnamefont{S.}~\bibnamefont{Kumano}},
  \bibinfo{author}{\bibfnamefont{J.-C.} \bibnamefont{Peng}},
  \bibinfo{author}{\bibfnamefont{S.}~\bibnamefont{Sawada}}, \bibnamefont{and}
  \bibinfo{author}{\bibfnamefont{K.}~\bibnamefont{Tanaka}},
  \bibinfo{journal}{Phys. Rev.} \textbf{\bibinfo{volume}{D93}},
  \bibinfo{pages}{114034} (\bibinfo{year}{2016}), \eprint{1605.00364}.

\bibitem[{\citenamefont{Lepage and Brodsky}(1979)}]{Lepage:1979zb}
\bibinfo{author}{\bibfnamefont{G.~P.} \bibnamefont{Lepage}} \bibnamefont{and}
  \bibinfo{author}{\bibfnamefont{S.~J.} \bibnamefont{Brodsky}},
  \bibinfo{journal}{Phys. Lett.} \textbf{\bibinfo{volume}{B87}},
  \bibinfo{pages}{359} (\bibinfo{year}{1979}).

\bibitem[{\citenamefont{Chernyak and Zhitnitsky}(1982)}]{Chernyak:1981zz}
\bibinfo{author}{\bibfnamefont{V.~L.} \bibnamefont{Chernyak}} \bibnamefont{and}
  \bibinfo{author}{\bibfnamefont{A.~R.} \bibnamefont{Zhitnitsky}},
  \bibinfo{journal}{Nucl. Phys.} \textbf{\bibinfo{volume}{B201}},
  \bibinfo{pages}{492} (\bibinfo{year}{1982}), \bibinfo{note}{[Erratum: Nucl.
  Phys.B214,547(1983)]}.

\bibitem[{\citenamefont{Ruiz~Arriola and
  Broniowski}(2006)}]{RuizArriola:2006jge}
\bibinfo{author}{\bibfnamefont{E.}~\bibnamefont{Ruiz~Arriola}}
  \bibnamefont{and}
  \bibinfo{author}{\bibfnamefont{W.}~\bibnamefont{Broniowski}},
  \bibinfo{journal}{Phys. Rev.} \textbf{\bibinfo{volume}{D74}},
  \bibinfo{pages}{034008} (\bibinfo{year}{2006}), \eprint{hep-ph/0605318}.

\bibitem[{\citenamefont{Radyushkin}(1994)}]{Radyushkin:1994xv}
\bibinfo{author}{\bibfnamefont{A.~V.} \bibnamefont{Radyushkin}}, in
  \emph{\bibinfo{booktitle}{{Workshop on Continuous Advances in QCD
  Minneapolis, Minnesota, February 18-20, 1994}}} (\bibinfo{year}{1994}), pp.
  \bibinfo{pages}{238--248}, \eprint{hep-ph/9406237}.

\bibitem[{\citenamefont{Ruiz~Arriola and
  Broniowski}(2002)}]{RuizArriola:2002bp}
\bibinfo{author}{\bibfnamefont{E.}~\bibnamefont{Ruiz~Arriola}}
  \bibnamefont{and}
  \bibinfo{author}{\bibfnamefont{W.}~\bibnamefont{Broniowski}},
  \bibinfo{journal}{Phys. Rev.} \textbf{\bibinfo{volume}{D66}},
  \bibinfo{pages}{094016} (\bibinfo{year}{2002}), \eprint{hep-ph/0207266}.

\bibitem[{\citenamefont{Chang et~al.}(2013)\citenamefont{Chang, Cloet,
  Cobos-Martinez, Roberts, Schmidt, and Tandy}}]{Chang:2013pq}
\bibinfo{author}{\bibfnamefont{L.}~\bibnamefont{Chang}},
  \bibinfo{author}{\bibfnamefont{I.~C.} \bibnamefont{Cloet}},
  \bibinfo{author}{\bibfnamefont{J.~J.} \bibnamefont{Cobos-Martinez}},
  \bibinfo{author}{\bibfnamefont{C.~D.} \bibnamefont{Roberts}},
  \bibinfo{author}{\bibfnamefont{S.~M.} \bibnamefont{Schmidt}},
  \bibnamefont{and} \bibinfo{author}{\bibfnamefont{P.~C.} \bibnamefont{Tandy}},
  \bibinfo{journal}{Phys. Rev. Lett.} \textbf{\bibinfo{volume}{110}},
  \bibinfo{pages}{132001} (\bibinfo{year}{2013}), \eprint{1301.0324}.

\bibitem[{\citenamefont{Agaev et~al.}(2012)\citenamefont{Agaev, Braun, Offen,
  and Porkert}}]{Agaev:2012tm}
\bibinfo{author}{\bibfnamefont{S.~S.} \bibnamefont{Agaev}},
  \bibinfo{author}{\bibfnamefont{V.~M.} \bibnamefont{Braun}},
  \bibinfo{author}{\bibfnamefont{N.}~\bibnamefont{Offen}}, \bibnamefont{and}
  \bibinfo{author}{\bibfnamefont{F.~A.} \bibnamefont{Porkert}},
  \bibinfo{journal}{Phys. Rev.} \textbf{\bibinfo{volume}{D86}},
  \bibinfo{pages}{077504} (\bibinfo{year}{2012}), \eprint{1206.3968}.

\bibitem[{\citenamefont{Braun et~al.}(2015)\citenamefont{Braun, Collins,
  Göckeler, Pérez-Rubio, Schäfer, Schiel, and Sternbeck}}]{Braun:2015axa}
\bibinfo{author}{\bibfnamefont{V.~M.} \bibnamefont{Braun}},
  \bibinfo{author}{\bibfnamefont{S.}~\bibnamefont{Collins}},
  \bibinfo{author}{\bibfnamefont{M.}~\bibnamefont{Göckeler}},
  \bibinfo{author}{\bibfnamefont{P.}~\bibnamefont{Pérez-Rubio}},
  \bibinfo{author}{\bibfnamefont{A.}~\bibnamefont{Schäfer}},
  \bibinfo{author}{\bibfnamefont{R.~W.} \bibnamefont{Schiel}},
  \bibnamefont{and}
  \bibinfo{author}{\bibfnamefont{A.}~\bibnamefont{Sternbeck}},
  \bibinfo{journal}{Phys. Rev.} \textbf{\bibinfo{volume}{D92}},
  \bibinfo{pages}{014504} (\bibinfo{year}{2015}), \eprint{1503.03656}.

\bibitem[{\citenamefont{Arthur et~al.}(2011)\citenamefont{Arthur, Boyle,
  Brommel, Donnellan, Flynn, Juttner, Rae, and Sachrajda}}]{Arthur:2010xf}
\bibinfo{author}{\bibfnamefont{R.}~\bibnamefont{Arthur}},
  \bibinfo{author}{\bibfnamefont{P.~A.} \bibnamefont{Boyle}},
  \bibinfo{author}{\bibfnamefont{D.}~\bibnamefont{Brommel}},
  \bibinfo{author}{\bibfnamefont{M.~A.} \bibnamefont{Donnellan}},
  \bibinfo{author}{\bibfnamefont{J.~M.} \bibnamefont{Flynn}},
  \bibinfo{author}{\bibfnamefont{A.}~\bibnamefont{Juttner}},
  \bibinfo{author}{\bibfnamefont{T.~D.} \bibnamefont{Rae}}, \bibnamefont{and}
  \bibinfo{author}{\bibfnamefont{C.~T.~C.} \bibnamefont{Sachrajda}},
  \bibinfo{journal}{Phys. Rev.} \textbf{\bibinfo{volume}{D83}},
  \bibinfo{pages}{074505} (\bibinfo{year}{2011}), \eprint{1011.5906}.

\bibitem[{\citenamefont{Braun et~al.}(2006)}]{Braun:2006dg}
\bibinfo{author}{\bibfnamefont{V.~M.} \bibnamefont{Braun}}
  \bibnamefont{et~al.}, \bibinfo{journal}{Phys. Rev.}
  \textbf{\bibinfo{volume}{D74}}, \bibinfo{pages}{074501}
  (\bibinfo{year}{2006}), \eprint{hep-lat/0606012}.

\bibitem[{\citenamefont{Daniel et~al.}(1991)\citenamefont{Daniel, Gupta, and
  Richards}}]{Daniel:1990ah}
\bibinfo{author}{\bibfnamefont{D.}~\bibnamefont{Daniel}},
  \bibinfo{author}{\bibfnamefont{R.}~\bibnamefont{Gupta}}, \bibnamefont{and}
  \bibinfo{author}{\bibfnamefont{D.~G.} \bibnamefont{Richards}},
  \bibinfo{journal}{Phys. Rev.} \textbf{\bibinfo{volume}{D43}},
  \bibinfo{pages}{3715} (\bibinfo{year}{1991}).

\bibitem[{\citenamefont{Martinelli and Sachrajda}(1987)}]{Martinelli:1987si}
\bibinfo{author}{\bibfnamefont{G.}~\bibnamefont{Martinelli}} \bibnamefont{and}
  \bibinfo{author}{\bibfnamefont{C.~T.} \bibnamefont{Sachrajda}},
  \bibinfo{journal}{Phys. Lett.} \textbf{\bibinfo{volume}{B190}},
  \bibinfo{pages}{151} (\bibinfo{year}{1987}).

\bibitem[{\citenamefont{Cloët et~al.}(2013)\citenamefont{Cloët, Chang,
  Roberts, Schmidt, and Tandy}}]{Cloet:2013tta}
\bibinfo{author}{\bibfnamefont{I.~C.} \bibnamefont{Cloët}},
  \bibinfo{author}{\bibfnamefont{L.}~\bibnamefont{Chang}},
  \bibinfo{author}{\bibfnamefont{C.~D.} \bibnamefont{Roberts}},
  \bibinfo{author}{\bibfnamefont{S.~M.} \bibnamefont{Schmidt}},
  \bibnamefont{and} \bibinfo{author}{\bibfnamefont{P.~C.} \bibnamefont{Tandy}},
  \bibinfo{journal}{Phys. Rev. Lett.} \textbf{\bibinfo{volume}{111}},
  \bibinfo{pages}{092001} (\bibinfo{year}{2013}), \eprint{1306.2645}.

\bibitem[{\citenamefont{Davoudi and Savage}(2012)}]{Davoudi:2012ya}
\bibinfo{author}{\bibfnamefont{Z.}~\bibnamefont{Davoudi}} \bibnamefont{and}
  \bibinfo{author}{\bibfnamefont{M.~J.} \bibnamefont{Savage}},
  \bibinfo{journal}{Phys. Rev.} \textbf{\bibinfo{volume}{D86}},
  \bibinfo{pages}{054505} (\bibinfo{year}{2012}), \eprint{1204.4146}.

\bibitem[{\citenamefont{Detmold and Lin}(2006)}]{Detmold:2005gg}
\bibinfo{author}{\bibfnamefont{W.}~\bibnamefont{Detmold}} \bibnamefont{and}
  \bibinfo{author}{\bibfnamefont{C.~J.~D.} \bibnamefont{Lin}},
  \bibinfo{journal}{Phys. Rev.} \textbf{\bibinfo{volume}{D73}},
  \bibinfo{pages}{014501} (\bibinfo{year}{2006}), \eprint{hep-lat/0507007}.

\bibitem[{\citenamefont{Monahan and Orginos}(2016)}]{Monahan:2016bvm}
\bibinfo{author}{\bibfnamefont{C.}~\bibnamefont{Monahan}} \bibnamefont{and}
  \bibinfo{author}{\bibfnamefont{K.}~\bibnamefont{Orginos}}
  (\bibinfo{year}{2016}), \eprint{1612.01584}.

\bibitem[{\citenamefont{Ji}(2013)}]{Ji:2013dva}
\bibinfo{author}{\bibfnamefont{X.}~\bibnamefont{Ji}}, \bibinfo{journal}{Phys.
  Rev. Lett.} \textbf{\bibinfo{volume}{110}}, \bibinfo{pages}{262002}
  (\bibinfo{year}{2013}), \eprint{1305.1539}.

\bibitem[{\citenamefont{Ji}(2014)}]{Ji:2014gla}
\bibinfo{author}{\bibfnamefont{X.}~\bibnamefont{Ji}}, \bibinfo{journal}{Sci.
  China Phys. Mech. Astron.} \textbf{\bibinfo{volume}{57}},
  \bibinfo{pages}{1407} (\bibinfo{year}{2014}), \eprint{1404.6680}.

\bibitem[{\citenamefont{Braun and Mueller}(2008)}]{Braun:2007wv}
\bibinfo{author}{\bibfnamefont{V.}~\bibnamefont{Braun}} \bibnamefont{and}
  \bibinfo{author}{\bibfnamefont{D.}~\bibnamefont{Mueller}},
  \bibinfo{journal}{Eur. Phys. J.} \textbf{\bibinfo{volume}{C55}},
  \bibinfo{pages}{349} (\bibinfo{year}{2008}), \eprint{0709.1348}.

\bibitem[{\citenamefont{Liu and Dong}(1994)}]{Liu:1993cv}
\bibinfo{author}{\bibfnamefont{K.-F.} \bibnamefont{Liu}} \bibnamefont{and}
  \bibinfo{author}{\bibfnamefont{S.-J.} \bibnamefont{Dong}},
  \bibinfo{journal}{Phys. Rev. Lett.} \textbf{\bibinfo{volume}{72}},
  \bibinfo{pages}{1790} (\bibinfo{year}{1994}), \eprint{hep-ph/9306299}.

\bibitem[{\citenamefont{Liu et~al.}(1999)\citenamefont{Liu, Dong, Draper,
  Leinweber, Sloan, Wilcox, and Woloshyn}}]{Liu:1998um}
\bibinfo{author}{\bibfnamefont{K.~F.} \bibnamefont{Liu}},
  \bibinfo{author}{\bibfnamefont{S.~J.} \bibnamefont{Dong}},
  \bibinfo{author}{\bibfnamefont{T.}~\bibnamefont{Draper}},
  \bibinfo{author}{\bibfnamefont{D.}~\bibnamefont{Leinweber}},
  \bibinfo{author}{\bibfnamefont{J.~H.} \bibnamefont{Sloan}},
  \bibinfo{author}{\bibfnamefont{W.}~\bibnamefont{Wilcox}}, \bibnamefont{and}
  \bibinfo{author}{\bibfnamefont{R.~M.} \bibnamefont{Woloshyn}},
  \bibinfo{journal}{Phys. Rev.} \textbf{\bibinfo{volume}{D59}},
  \bibinfo{pages}{112001} (\bibinfo{year}{1999}), \eprint{hep-ph/9806491}.

\bibitem[{\citenamefont{Liu}(2000)}]{Liu:1999ak}
\bibinfo{author}{\bibfnamefont{K.-F.} \bibnamefont{Liu}},
  \bibinfo{journal}{Phys. Rev.} \textbf{\bibinfo{volume}{D62}},
  \bibinfo{pages}{074501} (\bibinfo{year}{2000}), \eprint{hep-ph/9910306}.

\bibitem[{\citenamefont{Liu}(2016)}]{Liu:2016djw}
\bibinfo{author}{\bibfnamefont{K.-F.} \bibnamefont{Liu}},
  \bibinfo{journal}{PoS} \textbf{\bibinfo{volume}{LATTICE2015}},
  \bibinfo{pages}{115} (\bibinfo{year}{2016}), \eprint{1603.07352}.

\bibitem[{\citenamefont{Ma and Qiu}(2014)}]{Ma:2014jla}
\bibinfo{author}{\bibfnamefont{Y.-Q.} \bibnamefont{Ma}} \bibnamefont{and}
  \bibinfo{author}{\bibfnamefont{J.-W.} \bibnamefont{Qiu}}
  (\bibinfo{year}{2014}), \eprint{1404.6860}.

\bibitem[{\citenamefont{Xiong et~al.}(2014)\citenamefont{Xiong, Ji, Zhang, and
  Zhao}}]{Xiong:2013bka}
\bibinfo{author}{\bibfnamefont{X.}~\bibnamefont{Xiong}},
  \bibinfo{author}{\bibfnamefont{X.}~\bibnamefont{Ji}},
  \bibinfo{author}{\bibfnamefont{J.-H.} \bibnamefont{Zhang}}, \bibnamefont{and}
  \bibinfo{author}{\bibfnamefont{Y.}~\bibnamefont{Zhao}},
  \bibinfo{journal}{Phys. Rev.} \textbf{\bibinfo{volume}{D90}},
  \bibinfo{pages}{014051} (\bibinfo{year}{2014}), \eprint{1310.7471}.

\bibitem[{\citenamefont{Ji et~al.}(2015)\citenamefont{Ji, Schäfer, Xiong, and
  Zhang}}]{Ji:2015qla}
\bibinfo{author}{\bibfnamefont{X.}~\bibnamefont{Ji}},
  \bibinfo{author}{\bibfnamefont{A.}~\bibnamefont{Schäfer}},
  \bibinfo{author}{\bibfnamefont{X.}~\bibnamefont{Xiong}}, \bibnamefont{and}
  \bibinfo{author}{\bibfnamefont{J.-H.} \bibnamefont{Zhang}},
  \bibinfo{journal}{Phys. Rev.} \textbf{\bibinfo{volume}{D92}},
  \bibinfo{pages}{014039} (\bibinfo{year}{2015}), \eprint{1506.00248}.

\bibitem[{\citenamefont{Xiong and Zhang}(2015)}]{Xiong:2015nua}
\bibinfo{author}{\bibfnamefont{X.}~\bibnamefont{Xiong}} \bibnamefont{and}
  \bibinfo{author}{\bibfnamefont{J.-H.} \bibnamefont{Zhang}},
  \bibinfo{journal}{Phys. Rev.} \textbf{\bibinfo{volume}{D92}},
  \bibinfo{pages}{054037} (\bibinfo{year}{2015}), \eprint{1509.08016}.

\bibitem[{\citenamefont{Ji and Zhang}(2015)}]{Ji:2015jwa}
\bibinfo{author}{\bibfnamefont{X.}~\bibnamefont{Ji}} \bibnamefont{and}
  \bibinfo{author}{\bibfnamefont{J.-H.} \bibnamefont{Zhang}},
  \bibinfo{journal}{Phys. Rev.} \textbf{\bibinfo{volume}{D92}},
  \bibinfo{pages}{034006} (\bibinfo{year}{2015}), \eprint{1505.07699}.

\bibitem[{\citenamefont{Li}(2016)}]{Li:2016amo}
\bibinfo{author}{\bibfnamefont{H.-n.} \bibnamefont{Li}},
  \bibinfo{journal}{Phys. Rev.} \textbf{\bibinfo{volume}{D94}},
  \bibinfo{pages}{074036} (\bibinfo{year}{2016}), \eprint{1602.07575}.

\bibitem[{\citenamefont{Bali et~al.}(2016)\citenamefont{Bali, Lang, Musch, and
  Schäfer}}]{Bali:2016lva}
\bibinfo{author}{\bibfnamefont{G.~S.} \bibnamefont{Bali}},
  \bibinfo{author}{\bibfnamefont{B.}~\bibnamefont{Lang}},
  \bibinfo{author}{\bibfnamefont{B.~U.} \bibnamefont{Musch}}, \bibnamefont{and}
  \bibinfo{author}{\bibfnamefont{A.}~\bibnamefont{Schäfer}},
  \bibinfo{journal}{Phys. Rev.} \textbf{\bibinfo{volume}{D93}},
  \bibinfo{pages}{094515} (\bibinfo{year}{2016}), \eprint{1602.05525}.

\bibitem[{\citenamefont{Radyushkin}(2016)}]{Radyushkin:2016hsy}
\bibinfo{author}{\bibfnamefont{A.}~\bibnamefont{Radyushkin}}
  (\bibinfo{year}{2016}), \eprint{1612.05170}.

\bibitem[{\citenamefont{Radyushkin}(2017)}]{Radyushkin:2017gjd}
\bibinfo{author}{\bibfnamefont{A.~V.} \bibnamefont{Radyushkin}}
  (\bibinfo{year}{2017}), \eprint{1701.02688}.

\bibitem[{\citenamefont{Ishikawa et~al.}(2016)\citenamefont{Ishikawa, Ma, Qiu,
  and Yoshida}}]{Ishikawa:2016znu}
\bibinfo{author}{\bibfnamefont{T.}~\bibnamefont{Ishikawa}},
  \bibinfo{author}{\bibfnamefont{Y.-Q.} \bibnamefont{Ma}},
  \bibinfo{author}{\bibfnamefont{J.-W.} \bibnamefont{Qiu}}, \bibnamefont{and}
  \bibinfo{author}{\bibfnamefont{S.}~\bibnamefont{Yoshida}}
  (\bibinfo{year}{2016}), \eprint{1609.02018}.

\bibitem[{\citenamefont{Chen et~al.}(2016{\natexlab{a}})\citenamefont{Chen, Ji,
  and Zhang}}]{Chen:2016fxx}
\bibinfo{author}{\bibfnamefont{J.-W.} \bibnamefont{Chen}},
  \bibinfo{author}{\bibfnamefont{X.}~\bibnamefont{Ji}}, \bibnamefont{and}
  \bibinfo{author}{\bibfnamefont{J.-H.} \bibnamefont{Zhang}}
  (\bibinfo{year}{2016}{\natexlab{a}}), \eprint{1609.08102}.

\bibitem[{\citenamefont{Musch et~al.}(2011)\citenamefont{Musch, Hagler, Negele,
  and Schafer}}]{Musch:2010ka}
\bibinfo{author}{\bibfnamefont{B.~U.} \bibnamefont{Musch}},
  \bibinfo{author}{\bibfnamefont{P.}~\bibnamefont{Hagler}},
  \bibinfo{author}{\bibfnamefont{J.~W.} \bibnamefont{Negele}},
  \bibnamefont{and} \bibinfo{author}{\bibfnamefont{A.}~\bibnamefont{Schafer}},
  \bibinfo{journal}{Phys. Rev.} \textbf{\bibinfo{volume}{D83}},
  \bibinfo{pages}{094507} (\bibinfo{year}{2011}), \eprint{1011.1213}.

\bibitem[{\citenamefont{Lin}(2014{\natexlab{a}})}]{Lin:2014gaa}
\bibinfo{author}{\bibfnamefont{H.-W.} \bibnamefont{Lin}},
  \bibinfo{journal}{Int. J. Mod. Phys. Conf. Ser.}
  \textbf{\bibinfo{volume}{25}}, \bibinfo{pages}{1460039}
  (\bibinfo{year}{2014}{\natexlab{a}}).

\bibitem[{\citenamefont{Lin}(2014{\natexlab{b}})}]{Lin:2014yra}
\bibinfo{author}{\bibfnamefont{H.-W.} \bibnamefont{Lin}},
  \bibinfo{journal}{PoS} \textbf{\bibinfo{volume}{LATTICE2013}},
  \bibinfo{pages}{293} (\bibinfo{year}{2014}{\natexlab{b}}).

\bibitem[{\citenamefont{Lin et~al.}(2015)\citenamefont{Lin, Chen, Cohen, and
  Ji}}]{Lin:2014zya}
\bibinfo{author}{\bibfnamefont{H.-W.} \bibnamefont{Lin}},
  \bibinfo{author}{\bibfnamefont{J.-W.} \bibnamefont{Chen}},
  \bibinfo{author}{\bibfnamefont{S.~D.} \bibnamefont{Cohen}}, \bibnamefont{and}
  \bibinfo{author}{\bibfnamefont{X.}~\bibnamefont{Ji}}, \bibinfo{journal}{Phys.
  Rev.} \textbf{\bibinfo{volume}{D91}}, \bibinfo{pages}{054510}
  (\bibinfo{year}{2015}), \eprint{1402.1462}.

\bibitem[{\citenamefont{Alexandrou et~al.}(2015)\citenamefont{Alexandrou,
  Cichy, Drach, Garcia-Ramos, Hadjiyiannakou, Jansen, Steffens, and
  Wiese}}]{Alexandrou:2015rja}
\bibinfo{author}{\bibfnamefont{C.}~\bibnamefont{Alexandrou}},
  \bibinfo{author}{\bibfnamefont{K.}~\bibnamefont{Cichy}},
  \bibinfo{author}{\bibfnamefont{V.}~\bibnamefont{Drach}},
  \bibinfo{author}{\bibfnamefont{E.}~\bibnamefont{Garcia-Ramos}},
  \bibinfo{author}{\bibfnamefont{K.}~\bibnamefont{Hadjiyiannakou}},
  \bibinfo{author}{\bibfnamefont{K.}~\bibnamefont{Jansen}},
  \bibinfo{author}{\bibfnamefont{F.}~\bibnamefont{Steffens}}, \bibnamefont{and}
  \bibinfo{author}{\bibfnamefont{C.}~\bibnamefont{Wiese}},
  \bibinfo{journal}{Phys. Rev.} \textbf{\bibinfo{volume}{D92}},
  \bibinfo{pages}{014502} (\bibinfo{year}{2015}), \eprint{1504.07455}.

\bibitem[{\citenamefont{Chen et~al.}(2016{\natexlab{b}})\citenamefont{Chen,
  Cohen, Ji, Lin, and Zhang}}]{Chen:2016utp}
\bibinfo{author}{\bibfnamefont{J.-W.} \bibnamefont{Chen}},
  \bibinfo{author}{\bibfnamefont{S.~D.} \bibnamefont{Cohen}},
  \bibinfo{author}{\bibfnamefont{X.}~\bibnamefont{Ji}},
  \bibinfo{author}{\bibfnamefont{H.-W.} \bibnamefont{Lin}}, \bibnamefont{and}
  \bibinfo{author}{\bibfnamefont{J.-H.} \bibnamefont{Zhang}},
  \bibinfo{journal}{Nucl. Phys.} \textbf{\bibinfo{volume}{B911}},
  \bibinfo{pages}{246} (\bibinfo{year}{2016}{\natexlab{b}}),
  \eprint{1603.06664}.

\bibitem[{\citenamefont{Alexandrou et~al.}(2016)\citenamefont{Alexandrou,
  Cichy, Constantinou, Hadjiyiannakou, Jansen, Steffens, and
  Wiese}}]{Alexandrou:2016jqi}
\bibinfo{author}{\bibfnamefont{C.}~\bibnamefont{Alexandrou}},
  \bibinfo{author}{\bibfnamefont{K.}~\bibnamefont{Cichy}},
  \bibinfo{author}{\bibfnamefont{M.}~\bibnamefont{Constantinou}},
  \bibinfo{author}{\bibfnamefont{K.}~\bibnamefont{Hadjiyiannakou}},
  \bibinfo{author}{\bibfnamefont{K.}~\bibnamefont{Jansen}},
  \bibinfo{author}{\bibfnamefont{F.}~\bibnamefont{Steffens}}, \bibnamefont{and}
  \bibinfo{author}{\bibfnamefont{C.}~\bibnamefont{Wiese}}
  (\bibinfo{year}{2016}), \eprint{1610.03689}.

\bibitem[{\citenamefont{Adamczyk et~al.}(2014)}]{Adamczyk:2014xyw}
\bibinfo{author}{\bibfnamefont{L.}~\bibnamefont{Adamczyk}} \bibnamefont{et~al.}
  (\bibinfo{collaboration}{STAR}), \bibinfo{journal}{Phys. Rev. Lett.}
  \textbf{\bibinfo{volume}{113}}, \bibinfo{pages}{072301}
  (\bibinfo{year}{2014}), \eprint{1404.6880}.

\bibitem[{\citenamefont{Adare et~al.}(2016)}]{Adare:2015gsd}
\bibinfo{author}{\bibfnamefont{A.}~\bibnamefont{Adare}} \bibnamefont{et~al.}
  (\bibinfo{collaboration}{PHENIX}), \bibinfo{journal}{Phys. Rev.}
  \textbf{\bibinfo{volume}{D93}}, \bibinfo{pages}{051103}
  (\bibinfo{year}{2016}), \eprint{1504.07451}.

\bibitem[{\citenamefont{Follana et~al.}(2007)\citenamefont{Follana, Mason,
  Davies, Hornbostel, Lepage, Shigemitsu, Trottier, and Wong}}]{Follana:2006rc}
\bibinfo{author}{\bibfnamefont{E.}~\bibnamefont{Follana}},
  \bibinfo{author}{\bibfnamefont{Q.}~\bibnamefont{Mason}},
  \bibinfo{author}{\bibfnamefont{C.}~\bibnamefont{Davies}},
  \bibinfo{author}{\bibfnamefont{K.}~\bibnamefont{Hornbostel}},
  \bibinfo{author}{\bibfnamefont{G.~P.} \bibnamefont{Lepage}},
  \bibinfo{author}{\bibfnamefont{J.}~\bibnamefont{Shigemitsu}},
  \bibinfo{author}{\bibfnamefont{H.}~\bibnamefont{Trottier}}, \bibnamefont{and}
  \bibinfo{author}{\bibfnamefont{K.}~\bibnamefont{Wong}}
  (\bibinfo{collaboration}{HPQCD, UKQCD}), \bibinfo{journal}{Phys. Rev.}
  \textbf{\bibinfo{volume}{D75}}, \bibinfo{pages}{054502}
  (\bibinfo{year}{2007}), \eprint{hep-lat/0610092}.

\bibitem[{\citenamefont{Bazavov et~al.}(2013)}]{Bazavov:2012xda}
\bibinfo{author}{\bibfnamefont{A.}~\bibnamefont{Bazavov}} \bibnamefont{et~al.}
  (\bibinfo{collaboration}{MILC}), \bibinfo{journal}{Phys. Rev.}
  \textbf{\bibinfo{volume}{D87}}, \bibinfo{pages}{054505}
  (\bibinfo{year}{2013}), \eprint{1212.4768}.

\bibitem[{\citenamefont{Hasenfratz and Knechtli}(2001)}]{Hasenfratz:2001hp}
\bibinfo{author}{\bibfnamefont{A.}~\bibnamefont{Hasenfratz}} \bibnamefont{and}
  \bibinfo{author}{\bibfnamefont{F.}~\bibnamefont{Knechtli}},
  \bibinfo{journal}{Phys. Rev.} \textbf{\bibinfo{volume}{D64}},
  \bibinfo{pages}{034504} (\bibinfo{year}{2001}), \eprint{hep-lat/0103029}.

\bibitem[{\citenamefont{Bhattacharya et~al.}(2016)\citenamefont{Bhattacharya,
  Cirigliano, Cohen, Gupta, Lin, and Yoon}}]{Bhattacharya:2016zcn}
\bibinfo{author}{\bibfnamefont{T.}~\bibnamefont{Bhattacharya}},
  \bibinfo{author}{\bibfnamefont{V.}~\bibnamefont{Cirigliano}},
  \bibinfo{author}{\bibfnamefont{S.}~\bibnamefont{Cohen}},
  \bibinfo{author}{\bibfnamefont{R.}~\bibnamefont{Gupta}},
  \bibinfo{author}{\bibfnamefont{H.-W.} \bibnamefont{Lin}}, \bibnamefont{and}
  \bibinfo{author}{\bibfnamefont{B.}~\bibnamefont{Yoon}},
  \bibinfo{journal}{Phys. Rev.} \textbf{\bibinfo{volume}{D94}},
  \bibinfo{pages}{054508} (\bibinfo{year}{2016}), \eprint{1606.07049}.

\bibitem[{\citenamefont{Bhattacharya
  et~al.}(2015{\natexlab{a}})\citenamefont{Bhattacharya, Cirigliano, Cohen,
  Gupta, Joseph, Lin, and Yoon}}]{Bhattacharya:2015wna}
\bibinfo{author}{\bibfnamefont{T.}~\bibnamefont{Bhattacharya}},
  \bibinfo{author}{\bibfnamefont{V.}~\bibnamefont{Cirigliano}},
  \bibinfo{author}{\bibfnamefont{S.}~\bibnamefont{Cohen}},
  \bibinfo{author}{\bibfnamefont{R.}~\bibnamefont{Gupta}},
  \bibinfo{author}{\bibfnamefont{A.}~\bibnamefont{Joseph}},
  \bibinfo{author}{\bibfnamefont{H.-W.} \bibnamefont{Lin}}, \bibnamefont{and}
  \bibinfo{author}{\bibfnamefont{B.}~\bibnamefont{Yoon}}
  (\bibinfo{collaboration}{PNDME}), \bibinfo{journal}{Phys. Rev.}
  \textbf{\bibinfo{volume}{D92}}, \bibinfo{pages}{094511}
  (\bibinfo{year}{2015}{\natexlab{a}}), \eprint{1506.06411}.

\bibitem[{\citenamefont{Bhattacharya
  et~al.}(2015{\natexlab{b}})\citenamefont{Bhattacharya, Cirigliano, Gupta,
  Lin, and Yoon}}]{Bhattacharya:2015esa}
\bibinfo{author}{\bibfnamefont{T.}~\bibnamefont{Bhattacharya}},
  \bibinfo{author}{\bibfnamefont{V.}~\bibnamefont{Cirigliano}},
  \bibinfo{author}{\bibfnamefont{R.}~\bibnamefont{Gupta}},
  \bibinfo{author}{\bibfnamefont{H.-W.} \bibnamefont{Lin}}, \bibnamefont{and}
  \bibinfo{author}{\bibfnamefont{B.}~\bibnamefont{Yoon}},
  \bibinfo{journal}{Phys. Rev. Lett.} \textbf{\bibinfo{volume}{115}},
  \bibinfo{pages}{212002} (\bibinfo{year}{2015}{\natexlab{b}}),
  \eprint{1506.04196}.

\bibitem[{\citenamefont{Bhattacharya et~al.}(2014)\citenamefont{Bhattacharya,
  Cohen, Gupta, Joseph, Lin, and Yoon}}]{Bhattacharya:2013ehc}
\bibinfo{author}{\bibfnamefont{T.}~\bibnamefont{Bhattacharya}},
  \bibinfo{author}{\bibfnamefont{S.~D.} \bibnamefont{Cohen}},
  \bibinfo{author}{\bibfnamefont{R.}~\bibnamefont{Gupta}},
  \bibinfo{author}{\bibfnamefont{A.}~\bibnamefont{Joseph}},
  \bibinfo{author}{\bibfnamefont{H.-W.} \bibnamefont{Lin}}, \bibnamefont{and}
  \bibinfo{author}{\bibfnamefont{B.}~\bibnamefont{Yoon}},
  \bibinfo{journal}{Phys. Rev.} \textbf{\bibinfo{volume}{D89}},
  \bibinfo{pages}{094502} (\bibinfo{year}{2014}), \eprint{1306.5435}.

\bibitem[{\citenamefont{Chen and Stewart}(2004)}]{Chen:2003fp}
\bibinfo{author}{\bibfnamefont{J.-W.} \bibnamefont{Chen}} \bibnamefont{and}
  \bibinfo{author}{\bibfnamefont{I.~W.} \bibnamefont{Stewart}},
  \bibinfo{journal}{Phys. Rev. Lett.} \textbf{\bibinfo{volume}{92}},
  \bibinfo{pages}{202001} (\bibinfo{year}{2004}), \eprint{hep-ph/0311285}.

\bibitem[{\citenamefont{Edwards and Joo}(2005)}]{Edwards:2004sx}
\bibinfo{author}{\bibfnamefont{R.~G.} \bibnamefont{Edwards}} \bibnamefont{and}
  \bibinfo{author}{\bibfnamefont{B.}~\bibnamefont{Joo}}
  (\bibinfo{collaboration}{SciDAC, LHPC, UKQCD}), \bibinfo{journal}{Nucl. Phys.
  Proc. Suppl.} \textbf{\bibinfo{volume}{140}}, \bibinfo{pages}{832}
  (\bibinfo{year}{2005}), \bibinfo{note}{[,832(2004)]},
  \eprint{hep-lat/0409003}.

\bibitem[{\citenamefont{Braun and Manashov}(2011)}]{Braun:2011zr}
\bibinfo{author}{\bibfnamefont{V.~M.} \bibnamefont{Braun}} \bibnamefont{and}
  \bibinfo{author}{\bibfnamefont{A.~N.} \bibnamefont{Manashov}},
  \bibinfo{journal}{Phys. Rev. Lett.} \textbf{\bibinfo{volume}{107}},
  \bibinfo{pages}{202001} (\bibinfo{year}{2011}), \eprint{1108.2394}.

\bibitem[{\citenamefont{Braun and Manashov}(2012)}]{Braun:2011dg}
\bibinfo{author}{\bibfnamefont{V.~M.} \bibnamefont{Braun}} \bibnamefont{and}
  \bibinfo{author}{\bibfnamefont{A.~N.} \bibnamefont{Manashov}},
  \bibinfo{journal}{JHEP} \textbf{\bibinfo{volume}{01}}, \bibinfo{pages}{085}
  (\bibinfo{year}{2012}), \eprint{1111.6765}.

\end{thebibliography}
\end{document}